\documentclass[journal]{IEEEtran} 
\usepackage{graphicx}
\usepackage{epsfig}
\usepackage{epstopdf}
\usepackage{arydshln}
\usepackage{amsmath,amsthm,amsfonts,amssymb}
\usepackage{multirow}
\usepackage{cite}
\usepackage{verbatim}
\usepackage{tikz}
\usetikzlibrary{arrows}
\usetikzlibrary{backgrounds}
\usepackage{ellipsis}
\usepackage{xcolor}
\usepackage{xifthen}
\usepackage{cancel}
\theoremstyle{definition} 

\theoremstyle{definition} 
\theoremstyle{definition} 

\DeclareMathOperator{\F2}{\mathbb{F}_2}
\DeclareMathOperator{\sgn}{sgn}

\usetikzlibrary{shapes,arrows,intersections}
\usetikzlibrary{patterns,decorations}
\usetikzlibrary{mindmap}
\usetikzlibrary{calc}
\usepackage{ellipsis}
\usepackage{algorithm}
\usepackage{algpseudocode}
\usepackage{caption}

\newcommand{\fixme}[2]{\ifx&#2&{\leavevmode\color{red}#1}\else{\leavevmode\color{red}FIXME\{}#1{\leavevmode\color{red}\}}\footnote{{\leavevmode\color{red}#2}}\PackageWarning{Fixme}{#1: #2}\fi}

\newcommand{\newstuff}[2]{\ifx&#2&{\leavevmode\color{blue}#1}\else{\leavevmode\color{blue}FIXME\{}#1{\leavevmode\color{blue}\}}\footnote{{\leavevmode\color{blue}#2}}\PackageWarning{Newstuff}{#1: #2}\fi}

\begin{document}

\title{Memory Management in Successive-Cancellation based Decoders for Multi-Kernel Polar Codes}

\author{\IEEEauthorblockN{ Valerio Bioglio, Carlo Condo, Ingmar Land}\\
\IEEEauthorblockA{Mathematical and Algorithmic Sciences Lab\\ Huawei Technologies France SASU \\
Email: $\{$valerio.bioglio,carlo.condo,ingmar.land$\}$@huawei.com}} 

\maketitle

\begin{abstract}
Multi-kernel polar codes have recently been proposed to construct polar codes of lengths different from powers of two. 
Decoder implementations for multi-kernel polar codes need to account for this feature, that becomes critical in memory management.
We propose an efficient, generalized memory management framework for implementation of successive-cancellation decoding of multi-kernel polar codes. 
It can be used on many types of hardware architectures and different flavors of SC decoding algorithms. 
We illustrate the proposed solution for small kernel sizes, and give complexity estimates for various kernel combinations and code lengths.

\end{abstract}


\textbf{\textit{Topic Designation}--A. Communication Systems, 1. Modulation and Coding.}

\section{Introduction}

Polar codes \cite{polar} are a family of error correcting codes with capacity-achieving property over various classes of channels, providing excellent error rate performance for practical code lengths \cite{short_block}. 
The construction of polar codes is based on the polarization effect of the Kronecker powers of the binary $2 \times 2$ kernel matrix $T_2 = \begin{pmatrix} 1 & 0 \\ 1 & 1 \end{pmatrix}$. 
A major drawback of this construction is the restriction of achievable block lengths to powers of 2. 
Puncturing and shortening techniques can be used to adjust the code length, at the cost of a reduced bit polarization \cite{punct_paper}. 
To overcome this limitation, multi-kernel polar codes have been introduced in \cite{mk_arxiv}. 
By mixing binary kernels of different sizes in the construction of the code, these codes prove that many block lengths can be achieved while keeping the polarization effect. 

Many software and hardware implementations of polar code decoders have been proposed in literature. 
While software guarantees a higher degree of flexibility in terms of data structures, fast software decoders have to rely on efficient memory management \cite{legal_SW,Shen_SW}. 
The importance of smart memory usage is even more evident in hardware implementations, where memory accounts for the majority of area occupation and power consumption, and heavily impacts decoder speed \cite{SC_first,hashemi_JETCAS,Ercan_Allerton}. 
The memory structure first proposed in \cite{SC_semi-par} for purely binary polar codes, and widely adopted in SC-based decoders \cite{balatsoukas_SCL_HW}, relies on the observation that memory requirements decrease as the decoding stage increases. 
We show how this trend continues in multi-kernel polar codes, proposing an efficient memory structure for SC-based polar decoders, and providing functions for the evaluation of the overall memory requirements. 
This structure supports the decoding of codes constructed with any combination of kernel sizes, making it an ideal framework for multi-kernel decoder hardware implementations \cite{Coppolino_MK}. 

\section{Multi-Kernel Polar Codes}
\begin{figure}
\centering
\[ G_{12} = \left(
\begin{array}{ c c c:c c c|c c c:c c c}
 1 & 1 & 1 & 0 & 0 & 0 & 0 & 0 & 0 & 0 & 0 & 0 \\
 1 & 0 & 1 & 0 & 0 & 0 & 0 & 0 & 0 & 0 & 0 & 0 \\
 0 & 1 & 1 & 0 & 0 & 0 & 0 & 0 & 0 & 0 & 0 & 0 \\
 \hdashline
 1 & 1 & 1 & 1 & 1 & 1 & 0 & 0 & 0 & 0 & 0 & 0 \\
 1 & 0 & 1 & 1 & 0 & 1 & 0 & 0 & 0 & 0 & 0 & 0 \\
 0 & 1 & 1 & 0 & 1 & 1 & 0 & 0 & 0 & 0 & 0 & 0 \\
 \hline
 1 & 1 & 1 & 0 & 0 & 0 & 1 & 1 & 1 & 0 & 0 & 0 \\
 1 & 0 & 1 & 0 & 0 & 0 & 1 & 0 & 1 & 0 & 0 & 0 \\
 0 & 1 & 1 & 0 & 0 & 0 & 0 & 1 & 1 & 0 & 0 & 0 \\
 \hdashline
 1 & 1 & 1 & 1 & 1 & 1 & 1 & 1 & 1 & 1 & 1 & 1 \\
 1 & 0 & 1 & 1 & 0 & 1 & 1 & 0 & 1 & 1 & 0 & 1 \\
 0 & 1 & 1 & 0 & 1 & 1 & 0 & 1 & 1 & 0 & 1 & 1 
\end{array}
\right) \]
\caption{Transformation matrix $G_{12} = T_2 \otimes T_2 \otimes T_3$.}
\label{Fig:G_12_TM}
\end{figure}

Multi-kernel polar codes generalize the construction of polar codes by mixing binary kernels of different sizes. 
Similarly to polar codes, an $(N,K)$ multi-kernel polar code is completely defined by a $N \times N$ transformation matrix $G_N$ and a frozen set $\mathcal{F}$, with $|\mathcal{F}| = N-K$. 
Transformation matrix has the form 
\begin{equation}
  G_N = T_{p_1} \otimes T_{p_2} \cdots \otimes T_{p_s} ,
  \label{equ:GN-MK}
\end{equation}
where $T_{p_i}$ is a $p_i \times p_i$ binary matrix, $i=1,2,\ldots,s$, denoting a polarizing kernel of size $p_i$, and $N = p_1 \cdot p_2 \cdot \dots \cdot p_s$. 
Binary kernels of different sizes can be found in \cite{kernel_presman}. 
Transformation matrix $G_{12} = T_2 \otimes T_2 \otimes T_3$ is shown in Figure~\ref{Fig:G_12_TM}, where 
\begin{align}
\label{equ:T2_T3}
  T_2 & = \begin{pmatrix} 1 & 0 \\ 1 & 1 \end{pmatrix}, &
  T_3 & = \begin{pmatrix} 1 & 1 & 1 \\ 1 & 0 & 1 \\ 0 & 1 & 1 \end{pmatrix}  
\end{align}
and the recursive structure of the matrix is highlighted. 
The frozen set $\mathcal{F}$ indicates the $N-K$ bits to be frozen in the code construction, and can generally be designed according to bit reliabilities \cite{mk_arxiv} or minimum distance \cite{mk_dist_arxiv}. 
Finally, the encoder is defined by $x = u \cdot G_N$, mapping the input vector $u \in \F2^N$ to the codeword $x \in \F2^N$, where $u_i = 0$ for $i \in \mathcal{F}$, and $u_i$, $i \notin \mathcal{F}$, stores the information bits. 
We recall the set $\mathcal{I} = \mathcal{F}^c$ to be termed as information set. 

\begin{figure}
\centering
\includegraphics[width=\columnwidth]{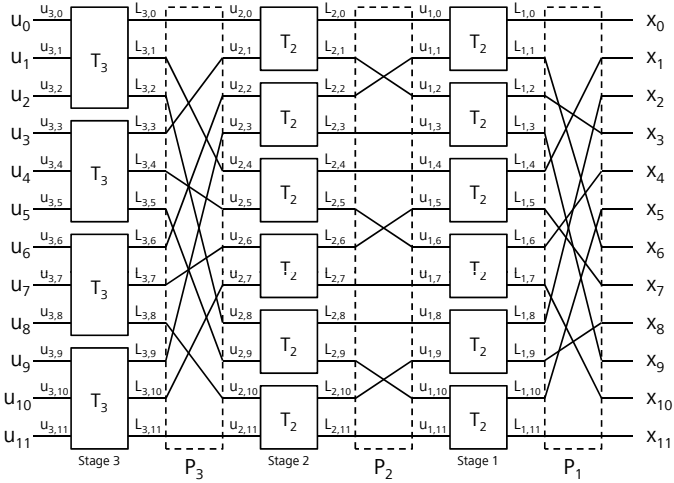}
\caption{Tanner graph defined by $G_{12} = T_2 \otimes T_2 \otimes T_3$.}
\label{Fig:G_12_TG}
\end{figure}

The structure of multi-kernel polar codes can be better understood through the Tanner graph of the code; this consists of various $p_i \times p_i$ blocks $B_{p_i}$, corresponding to the different $T_{p_i}$ kernels used in the construction of the transformation matrix, connecting input vector and codeword.
Each of the $s$ stages composing the graph is formed by $N_i = N/p_i$ kernel blocks $B_{p_i}$, performing the operations involving kernel $T_{p_i}$. 
Permutations $P_i$ between stages are described in \cite{mk_arxiv}; an example of Tanner graph for a $G_{12}$ is given in Figure~\ref{Fig:G_12_TG}.

Multi-kernel polar codes can be decoded through successive cancellation (SC) decoding on the Tanner graph of the code, where log-likelihood ratios (LLRs) \cite{radix} are passed from the right to the left, while partial sums (PSs) based on hard decisions on the decoded bits are passed from the left to the right. 
LLRs and PSs are calculated in the kernel blocks, depicted as
\begin{center}
  \begin{tikzpicture}[scale=0.2, line width=0.8pt]
	\draw   (0,0) rectangle (4,8) node[midway]{$B_p$};
	\draw[] (-2,7) node[left] {$u_0 , l_0$} -- (0,7) ;
	\draw[] (-2,5) node[left] {$u_1 , l_1$} -- (0,5) ;
	\draw[] (-0.5,3.5) node[left] {$\vdots$}    (0,3) ;
	\draw[] (-2,1) node[left] {$u_{p-1} , l_{p-1}$} -- (0,1) ;
	\draw[] (4,7)                     -- (6,7) node[right] {$x_0 , L_0$} ;
	\draw[] (4,5)                     -- (6,5) node[right] {$x_1 , L_1$} ;
	\draw[] (4,3)                        (4.5,3.5) node[right] {$\vdots$} ;
	\draw[] (4,1)                     -- (6,1) node[right] {$x_{p-1} , L_{p-1}$} ;
  \end{tikzpicture}
\end{center}
Blocks in the same column belong to the same stage and can perform decoding operations in parallel. 
Roughly speaking, $L_i$ and $l_i$ represent the LLRs of the partial sums $u_i$ and $x_i$ respectively. 
However, PSs are calculated on the basis of the previously decoded bit, hence they may not match with the connected LLRs. 
We indicate with $L_{i,(j-1)p_i},\dots,L_{i,j p_i - 1}$ and $u_{i,(j-1)p_i},\dots,u_{i,j p_i - 1}$ the LLRs and PSs input of the $j$-th block of stage $i$ respectively, with $j \leq N_i = N/p_i$. 
LLRs $L_{1,0},\dots,L_{1,N-1}$ correspond to channel LLRs, while $u_{s,0},\dots,u_{s,N-1}$ correspond to the decoded bits. 
An example of this labeling is given in Figure~\ref{Fig:G_12_TG}. 

Given the binary input vector $u = (u_0,u_1,\ldots,u_{p-1})$, corresponding to the partial sums calculated from the decoded bits, the output vector $x = (x_0,x_1,\ldots,x_{p-1})$ is calculated as $x = u \cdot T_p$. 
If we call $T_p^i$ the $i$-th column of the kernel matrix $T_p$, the update rule for the PSs can be written as
\begin{equation}
 x_i = u \cdot T_p^i. 
 \label{equ:g}
\end{equation} 
The vector $x$ corresponds to the partial sums calculated by the kernel $T_p$, that will be used as input for the LLRs calculations of other blocks. 
This update rule is performed from left to right, and can be used also for the encoding. 

Output LLRs $l_0, \dots, l_{p-1}$ are calculated sequentially using the input LLRs $L_0,\dots,L_{p-1}$ coming from the previous stage and the PSs corresponding to the previously decoded bit, i.e., 
\begin{equation}
 l_i = f_i^{p}( L_0 , \dots , L_{p-1} , u_0 , \dots, u_{i-1} ), 
 \label{equ:f}
\end{equation}  
with $l_0 = f_0^{p}( L_0 , \dots , L_{p-1} )$. 
This update is performed from the right to the left, and corresponds to the successive cancellation principle. 
Rules for the derivation of LLR update functions for arbitrary binary kernels can be found in \cite{marg_ker}. 

\section{Memory Management}
\begin{figure}
\centering
\includegraphics[width=0.98\columnwidth]{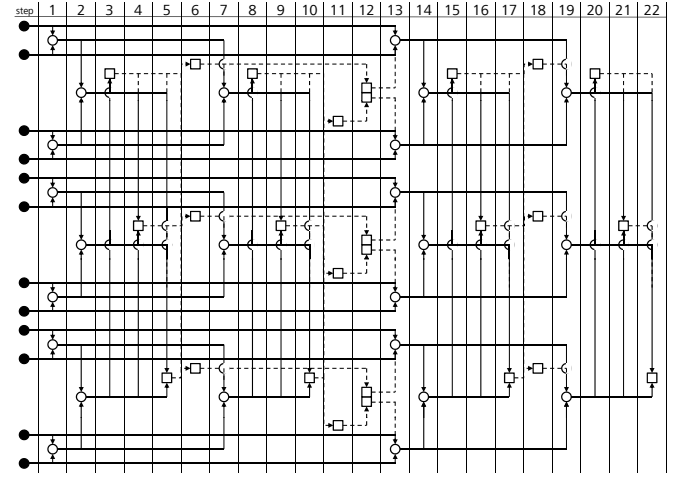}
\caption{Data flow graph of the SC decoder for the multi-kernel polar code defined by $G_{12} = T_2 \otimes T_2 \otimes T_3$. Circles represent LLR updates, squares represent PS updates.}
\label{Fig:G_12_flow}
\end{figure} 

Similarly to polar codes, it is possible to describe the SC decoding process of a multi-kernel polar code using a data flow graph, depicted in Figure~\ref{Fig:G_12_flow} for the code generated by $G_{12}$. 
The data flow graph represents memory dependencies arising during the decoding, where circles and squares represent memory needed to store LLRs and PSs, respectively, and black circles represent channel LLRs. 
In particular, circles and squares identify the need for new memory allocation, while horizontal lines determine the number of time steps for which the values need to be stored.
Thick lines represent LLR updates, while dotted lines identify operations involving partial sums, i.e. LLR updates when they merge with thick lines and PSs updates when they connect squares. 
The study of the data flow graph highlights the strong dependencies among data, along with a repetitive structure in the LLR update functions, and gives a precise order in the scheduling of the decoding operations. 
In general, the hardest constraint for the LLR update functions is given by the calculation of the necessary PSs.
The memory usage patterns observed in Figure~\ref{Fig:G_12_flow} can be found for code of any length, and can be exploited to develop a memory management framework as follows. 

\subsection{Memory Structure}

\begin{figure}
\centering
\includegraphics[width=\columnwidth]{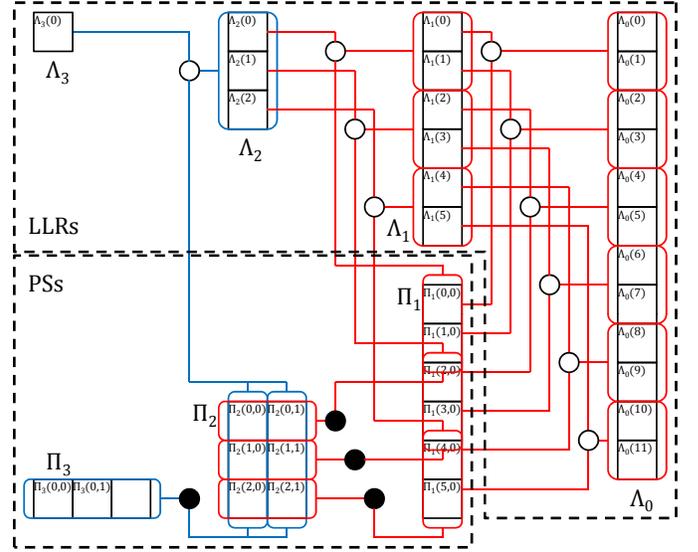}
\caption{Memory structure for $G_{12} = T_2 \otimes T_2 \otimes T_3$. White dots represent LLR updates, black dots represent PS updates. Red lines represent $B_2$ blocks, blue lines represent $B_3$ blocks.}
\label{Fig:G_12_MS}
\end{figure}

The memory structures for a generic SC decoder for the multi-kernel polar code defined by $G_{12}$ is presented in Figure~\ref{Fig:G_12_MS}, along with memory dependencies. 
We call $\Lambda$, $\Pi$ and $\Upsilon$ the data structures used to respectively store LLRs, PSs and decoded bits. 
We define as $Q$ the number of bits assigned to the representation of each internal LLR, while a partial sum and a decoded bit are, by definition, single-bit values. 
The proposed memory structure relies on the observation made in in \cite{SC_semi-par} that memory requirements for polar codes decoding decrease as the stage index increases; we show that this phenomenon can be extended to multi-kernel polar codes.  

The memory structure of multi-kernel polar codes decoder depends on the order of the kernels defining the transformation matrix $G_N = T_{p_1} \otimes \dots \otimes T_{p_s}$, where $s$ is the number of stages, i.e., the number of factors in the Kronecker product. 
%
LLRs can be stored in $s+1$ $Q$-bits vectors $\Lambda_0,\dots,\Lambda_s$ of different lengths, i.e. with a different number of elements. 
The length of vector $\Lambda_s$ is always 1, and stores the LLR of the currently decoded bit. 
The length of vector $\Lambda_i$ is given by the product of the last $s-i$ kernel sizes, i.e., $\Lambda_i$ has $p_{i+1} \cdot \dots \cdot p_s$ entries. 
PSs are stored in $s$ binary matrices  $\Pi_1,\dots,\Pi_s$ of different width and depth, depending on the decoding stage. 
The width of $\Pi_i$ is given by the size of kernel $p_i$, while its depth is given by the product of the last $s-i$ kernel sizes, i.e., is given by $p_{i+1} \cdot \dots \cdot p_s$, similarly to LLR vectors. 
The first matrix $\Pi_s$ is an exception, since it has width $p_s-1$. 
This is due to the fact that the last column of $\Pi_s$ would be updated during the PS update phase of the decoding of the last bit. 
Since the PS update is executed right after the bit estimation, we skip this last PS update and we do not need to store the last column of $\Pi_s$.
Finally, the decoded bits are stored in the binary vector $\Upsilon$ of length $N$.

\subsection{Memory Update}

\begin{algorithm}[htb]
\caption{SC Algorithm} \label{algo}
\begin{algorithmic}[1]
\State $\text{Initialize } \Lambda, \Pi, \Upsilon $ 
\For{$i = 0 \dots N-1$}
   \State $\text{LLR update}$
   \State $u_i \text{ calculation}$
   \State $\text{PS update}$ 
\EndFor
\State\Return $u$
\end{algorithmic}
\end{algorithm}
%
Algorithm \ref{algo} depicts the logical flow of operations required by SC decoding. In this Section, we follow its schedule and first describe the update operations for the LLRs, then for the decoded bits $u$, and finally for the PSs. 
The memory update operations are performed by the kernel block. 
LLR vector $\Lambda_0$ is initially filled with the $N$ LLRs extracted from the received symbols, while the rest of the memory is initialized to zero; we recall that the LLRs have to be permuted according to $P_1$ before the insertion in $\Lambda_0$. 
According to the SC algorithm, bits are decoded sequentially, hence some of the memory structures are updated at every bit estimation; this update process is illustrated for the decoding of generic input bit $u_i$.

\textbf{LLR update: } \\
The update function \eqref{equ:f} to be used in this phase depends on the index $i$ of the decoded bit $u_i$, and it is selected using the mixed radix representation of $i$ based on the kernels composing the transformation matrix of the code. 

In a base $p$ radix system, positive integers are represented as a finite sequence of digits smaller than $p$.  
A mixed radix system is a non-standard positional numeral system, generalizing classic radix system, in which the numerical base depends on the digit position. 
A well known example of a mixed radix numeral system is the one used to measure time in hours, minutes and seconds.
We use the sequence $\langle p_1, \ldots p_s \rangle$ of the sizes of the kernels constructing the transformation matrix as the base of a finite mixed radix system representing the index $i$ of the decoded bit $u_i$. 
According to this representation, any integer $i < N$ can be expressed as a vector of $s$ digits $i = b_1^{(i)} \dots b_s^{(i)}$, with $0 \leq b_j^{(i)} < p_j$ and
\begin{equation}
i = b_s^{(i)} + \sum_{j=1}^{s-1} b_j^{(i)} \cdot (p_{j+1} \cdot \dots \cdot p_s). 
\end{equation} 
The mixed radix representation of the decoded bits indices for $G_{12} = T_2 \otimes T_2 \otimes T_3$ is given by:  
%
\begin{center}
\begin{tabular}{ c|c c c c c c c c c c c c c}
 $i$   & 0 & 1 & 2 & 3 & 4 & 5 & 6 & 7 & 8 & 9 & 10 & 11 \\
 \hline
 $b_1^{(i)}$ & 0 & 0 & 0 & 0 & 0 & 0 & 1 & 1 & 1 & 1 & 1  & 1  \\
 $b_2^{(i)}$ & 0 & 0 & 0 & 1 & 1 & 1 & 0 & 0 & 0 & 1 & 1  & 1  \\
 $b_3^{(i)}$ & 0 & 1 & 2 & 0 & 1 & 2 & 0 & 1 & 2 & 0 & 1  & 2  \\
\end{tabular}
\end{center}

The update of LLR vectors proceeds from right to left in Figure~\ref{Fig:G_12_MS}. 
Starting from $\Lambda_1$, all the vectors are updated using the previous LLR vector and the present PS matrix as input; in general, vector $\Lambda_j$ is updated using vector $\Lambda_{j-1}$ and the partial sums stored in $\Pi_j$. 
The LLR update rule to be used in the update of $\Lambda_j$ is selected using the mixed radix representation of $i$, and more precisely the LLR update rule $f_{b_j^{(i)}}^{p_j}$ is used. 
This method is an extension of the method proposed in \cite{SC_semi-par} for the scheduling of $f$ and $g$ functions in the decoding of polar codes. 
Each entry $\Lambda_j(k)$ of the LLR vector is calculated as 
\begin{equation}
\begin{array}{l}   
\Lambda_j(k) = f_{b_j^{(i)}}^{p_j}( \Lambda_{j-1}(k \cdot p_j) , \dots , \Lambda_{j-1}((k+1) \cdot p_j - 1) ,\\
\qquad \qquad \Pi_j(k,0) , \dots, \Pi_j(k,b_j^{(i)}) ). 
\end{array}
\end{equation}
The update operations of a vector $\Lambda_j$ can be run in parallel to reduce latency using up to $p_{j+1} \cdot \dots \cdot p_s$ kernel blocks. 

Using the proposed LLR update algorithm, $s$ LLR vectors, i.e. from $\Lambda_1$ to $\Lambda_s$, are updated for every decoded bit $u_i$. 
However, the data flow presented in Figure~\ref{Fig:G_12_flow} shows that the number of vectors to be updated actually depends on the index $i$ of the decoded bit. 
A closer look to the mixed radix representation table suggests the reason of this scheduling: in fact, the mixed radix representations of two consecutive numbers differ only on the right of the position of the rightmost nonzero element of the second number. 
In practice, given $i-1 = b_1^{(i-1)} \dots b_s^{(i-1)}$, if it exists an index $z$ such that $b_z^{(i)} \neq 0$ and $b_j^{(i)} = 0$ for all $j > z$, we have that $b_j^{(i)} = b_j^{(i-1)}$ for all $j < z$. 
As a consequence, to decode the bit $u_i$ it is not necessary to update the vectors $\Lambda_j$ with indices $j < z$, and the update can be run starting from $\Lambda_z$. 
Of course, for the case $i=0$ all the vectors have to be updated. 
This acceleration technique is a generalization of the one proposed in \cite{radix} for polar codes, and halves the number of vectors updates. 

This property allows a further simplification of the LLR update algorithm. 
We have seen that the LLR update is run starting from $\Lambda_z$ with $z$ such that $b_z^{(i)} \neq 0$ and $b_j^{(i)} = 0$ for all $j > z$. 
This means that the vector $\Lambda_z$ is updated using function $f_{b_z^{(i)}}^{p_j}$, while all the other vectors are updated using a function of the form $f_{0}^{p_j}$. 
This means that it is only necessary to find the subscript of the first LLR update function, while the other ones all have subscript 0.

\begin{algorithm}[htb]
\caption{LLR update} \label{algo_LLR}
\begin{algorithmic}[1]
\State $n = 1 $ 
\If{i = 0}
   \State break
\Else
   \For{$z = s \dots 1$}
      \If{$i \mod p_z \neq 0$}
         \State break
      \EndIf
      \State $i = \frac{i}{p_z}$
      \State $n = n \cdot p_z$
   \EndFor
\EndIf
\State $b = i \mod p_z$
\For{$k = 0 \dots n-1$}
   \State $\Lambda_z(k)= f_{b}^{p_z}( \Lambda_{z-1}(k \cdot p_z) , \dots , \Lambda_{z-1}((k+1) \cdot p_z - 1) , \Pi_z(k,0) , \dots, \Pi_z(k,b) )$
\EndFor
\For{$j = z+1 \dots z$}
   \State $n = n \cdot p_j$
   \For{$k = 0 \dots n-1$}
      \State $\Lambda_j(k)= f_{0}^{p_j}( \Lambda_{j-1}(k \cdot p_j) , \dots , \Lambda_{j-1}((k+1) \cdot p_j - 1) )$
   \EndFor
\EndFor
\end{algorithmic}
\end{algorithm}

\textbf{$u_i$ estimation: } \\
If $i \in \mathcal{F}$, i.e. it belongs to the frozen set, its value is known to be zero, hence $\Upsilon(i) = 0$. 
Otherwise, i.e. if $i \notin \mathcal{F}$, the value decoded bit is decided by hard decision on its LLR. 
After the LLR update phase, the LLR of the bit $u_i$ will be copied in $\Lambda_s$ as explained in next paragraph. 
In our implementation, negative LLRs represent the bit 1, while positive LLRs represent the bit 0. 
Through hard decision, we set $\Upsilon(i) = \frac{\sgn(\Lambda_s(0))+1}{2}$. 
To sum up, we have that
\begin{equation}
\Upsilon(i) = 
\left\{
\begin{array}{l r}   
  0 & \text{if } i \in \mathcal{F}   \\
  \frac{\sgn(\Lambda_s(0))+1}{2} & \text{if } i \notin \mathcal{F}
\end{array} 
\right.
\end{equation}

\begin{algorithm}[htb]
\caption{$u_i$ calculation} \label{algo_U}
\begin{algorithmic}[1]
\If{$i \in \mathcal{F}$}
   \State $\Upsilon(i) = 0$
\Else
   \State $\Upsilon(i) = \frac{\sgn(\Lambda_s(0))+1}{2}$
\EndIf
\end{algorithmic}
\end{algorithm}

\textbf{PS update: } \\
PS matrices are updated in decreasing order starting from $\Pi_s$. 
Inside each matrix, the entries update is performed per columns, in increasing order starting from the first column. 
When the last column of a matrix is filled, a column of the next matrix is updated. 
Similarly to LLRs, the update function depends on the mixed radix representation of index $i$; in particular, the number of matrices to be updated is given by the number of consecutive digits of the mixed radix representation of $i$ with the highest symbol admitted by the radix, counting from the last digit. 

Update always starts from the last PS matrix $\Pi_s$, that is a row vector of width $p_s$. 
The value of the decoded bit $u_i$ is copied in the column $b_s^{(i)}$ of the matrix, i.e., $\Pi_s(0,b_s^{(i)}) = \Upsilon(i)$. 
When $b_s=p_s-1$, the last column of the matrix has been filled, and the column $b_{s-1}^{(i)}$ of the matrix $\Pi_{s-1}$ is updated, otherwise the update process ends. 
In general, if $b_j^{(i)} = p_j-1$ for all $j > z$ and $b_z^{(i)} < p_z-1$, the matrices $\Pi_s,\Pi_{s-1},\dots,\Pi_z$ are going to be updated. 
When the last column of matrix $\Pi_j$ is filled, i.e. when $b_j^{(i)} = p_j-1$, then the column $b_{j-1}$ of matrix $\Pi_{j-1}$ has to be updated. 
In this case, each row of $\Pi_j$ is used to update the column $b_{j-1}^{(i)}$ of $\Pi_{j-1}$ as $[\Pi_{j-1}(k \cdot p_j,b_{j-1}^{(i)}),\dots,\Pi_{j-1}((k+1)\cdot p_j - 1,b_{j-1}^{(i)})] = [\Pi_j(k,0),\dots,\Pi_j(k,p_j-1)] \cdot T_{p_j}$ for $k = 0,\dots,p_{j+1} \cdot \dots \cdot p_s - 1$. 
If we call $T_p^k$ the vector formed by the $k$-th column of the kernel matrix $T_p$, the update rule for the PSs can be rewritten as
\begin{equation}
\Pi_j(k,b_j^{(i)}) = \Pi_j \left( \left\lfloor \frac{k}{p_{j-1}} \right\rfloor ,- \right) \cdot T_{p_j}^{c}
\end{equation}
for $k = 0,\dots,p_{j+1} \cdot \dots \cdot p_s$, where $\Pi_j(k,-)$ represents the $k$-th row of $\Pi_j$ and $c = (k \mod p_{j+1})+1$. 
As an exception, the PS update step is not executed for the last decoded bit $u_{N-1}$, since this phase would have been executed after the decoding of the last bit and it would be pointless. 

\begin{algorithm}[htb]
\caption{PS update} \label{algo_PS}
\begin{algorithmic}[1]
\State $n = 1$ 
\If{$i = N-1$}
   \State return
\EndIf
\For{$j = s-1 \dots 1$}
   \If{$i + 1 \mod p_{j+1} \neq 0$}
      \State return
   \EndIf
   \State $i = \frac{i+1}{p_{j+1}} - 1$
   \State $n = n \cdot p_{j+1}$
   \State $b = i \mod p_j$
   \For{$k = 0 \dots n-1$}
      \State $c = (k \mod p_{j+1})+1$
      \State $\Pi_j(k,b) = \Pi_{j+1} \left( \left\lfloor \frac{k}{p_{j-1}} \right\rfloor,- \right) \cdot T_{p_j}^{c}$
   \EndFor
\EndFor
\end{algorithmic}
\end{algorithm}

\section{Analysis and Conclusions}

The proposed memory structure allows to limit the memory requirement of a multi-kernel polar decoder. 
In fact, a na\"{\i}ve memory management of the SC decoder for a multi-kernel polar codes with transformation matrix $G_N=T_{p_1} \otimes \dots \otimes T_{p_s}$ requires to store 
all the LLRs and the PSs depicted in the Tanner graph of the code. 
As a consequence, $\mathcal{M}^{LLR}_{} = N \cdot (s+1)$ LLRs and $\mathcal{M}^{PS}_{} = N \cdot s$ PSs, with $N = p_1 \cdot \dots \cdot p_s$, have to be stored, with space complexity $O(s N)$. 
The memory requirement is hence linearly dependent on both the code length $N$ and the number of kernels $s$. 

In the proposed memory structure, every LLR vector $\Lambda_i$ with $i \leq 1$ stores $\frac{N}{p_1 \cdot \dots \cdot p_{i} }$ LLRs, while the first vector $\Lambda_0$ stores the $N$ LLRs derived from the received signals. 
In total, for the proposed memory framework
\begin{equation}
\begin{array}{r l}
  \mathcal{M}^{LLR}_{prop} & = N + \frac{N}{p_1} + \frac{N}{p_1 \cdot p_2} + \dots + 1 = \\
   & = \left( \dots \left( \left( p_1 + 1 \right) \cdot p_2 + 1 \right) \cdot \dots \right) \cdot p_s + 1
\end{array}
\end{equation}
LLRs have to be stored. 
Similarly, every PS matrix $\Pi_i$ with $i>1$ stores $\frac{N}{p_1 \cdot \dots \cdot p_{i} } \cdot p_i = \frac{N}{p_1 \cdot \dots \cdot p_{i-1} }$ partial sums, while $\Pi_1$ stores $\frac{N}{p_1} \cdot (p_s - 1)$ PSs. 
Then, the total number of PSs is 
\begin{equation}
\begin{array}{r l}
  \mathcal{M}^{PS}_{prop} & = \frac{N}{p_1} \cdot (p_s - 1) + \frac{N}{p_1 \cdot p_2} + \dots + p_s =  \\
   & = \left( \dots \left( \left( p_1 \cdot p_2 + 1 \right) \cdot p_3 + 1 \right) \cdot \dots \right) \cdot p_s.
\end{array}
\end{equation}
By construction, we have that $\mathcal{M}^{PS}_{prop} \leq N \leq \mathcal{M}^{LLR}_{prop} < 2N$, hence the space complexity for both LLRs and PSs is reduced to $O(N)$. 
A comparison between the memory requirements for the proposed memory structure and the na\"{\i}ve one involving only kernels of sizes 2 and 3 is presented here: 
%
\begin{center}
\begin{tabular}{|c|c|c|c|c|c|}
\hline
N	& 12 & 72 & 144 & 384 & 972 \\
\hline
$\mathcal{M}^{LLR}_{prop}$ & 22 & 139 &  283 &  766 & 1822 \\
$\mathcal{M}^{LLR}_{}$ & 48 & 432 & 1008 & 3456 & 7776 \\
\hline
$\mathcal{M}^{PS}_{prop}$  & 15 & 102 &  210 &  573 & 1335 \\
$\mathcal{M}^{PS}_{}$  & 36 & 360 &  864 & 3072 & 6804 \\
\hline	 
\end{tabular}
\end{center}
The memory requirement reduction enabled by the proposed memory structure is remarkable. 
This proves that multi-kernel polar codes can be used as a valid alternative to punctured polar codes in terms of memory complexity. 
Given the similarities between polar codes and multi-kernel polar codes, it is straightforward to apply the proposed memory structure to list or simplified SC decoders. 
Finally, the proposed implementation can be easily transposed to hardware, reducing the complexity of an ASIC or FPGA dedicated architecture. 

\bibliographystyle{IEEEbib}
\bibliography{polar_codes_bib}
\end{document}